\documentclass[epj]{svjour}
\usepackage{graphicx}
\usepackage[numbers,sort&compress]{natbib}
\usepackage{natbib}

\begin{document}
\title{Conductance through analytic constrictions} 
\author{D.\ Koudela\inst{1,2} \and A.-M.\ Uimonen\inst{1,2} \and H.\ 
H\"akkinen\inst{1,2,3}}
\mail{daniela.koudela@gmx.de}
\institute{Nanoscience Center, P.O. Box 35,
            FIN-40014 University of Jyv\"askyl\"a, Finland
		\and
	    Department of Physics, P.O. Box 35, 
	    FIN-40014 University of Jyv\"askyl\"a, Finland
		\and
	    Department of Chemistry, P.O. Box 35,
            FIN-40014 University of Jyv\"askyl\"a, Finland}
\date{Received: date / Revised version: date}
\abstract{
We study the dependence of the intrinsic conductance
of a nanocontact on its shape by
using the recursion-transfer-matrix method.
Hour-glass, torus, and spherical shapes are defined
through analytic potentials, the latter two
serving as rough models for ring-like and spherical molecules, respectively.
The sensitivity of the conductance to geometric details is 
analyzed and discussed.
Strong resonance effects are found for a spherical contact weakly coupled
to electron reservoirs.
}
\maketitle
\section{Introduction}\label{intro}
% Why is it interesting to study the conductance of nanocontacts?\\
Nanocontacts are a research topic of intense current interest, motivated by
the drive to reduce the size of integrated circuits \cite{Agrait2003_81}.
Molecules, molecule-like species and nanoparticles have been considered as 
natural building blocks for interconnects.
Especially Benzene-dithiolate 
\cite{Andrews2006_174718,Basch2005_1668,Bauschlicher2004_427,Choi2007_155420,Delaney2004_036805,Derosa2001_471,Di-Ventra2000_979,Emberly1998_10911,Emberly2001_235412,Emberly2003_188301,Faleev2005_195422,Ke2004_15897,Ke2005_074704,Maiti2008_126,Reed1997_252,Stokbro2003_151,Thygesen2005_111,Toher2007_056801,Ulrich2006_2462,Varga2007_076804,Viljas2007_033403}
and C$_{60}$ 
\cite{Alavi2002_293,Bohler2007_125432,Joachim1995_2102,Joachim1997_353,Kaun2005_226801,Neel2007_065502,Ono2007_026804,Park2000_57,Sergueev2007_233418}
have been extensively studied.

Gaining detailed knowledge of bias-dependent quantum transport requires
high-level ab initio calculations on the coupling of the molecular orbitals
between the interconnect and the leads. However, an instructive view
of ballistic properties of the contact can be achived via considering a 
general quantum mechanical transmission problem through an analytically
defined constriction of various shapes. 
Here we calculated the conductance of a 
torus connected to jellium electrodes as a crude model for a 
Benzene-ring and the conductance of two concentric spheres 
which can be seen as a model for C$_{60}$. In addition we 
investigated if two nanoconstrictions in close contact interact
with each other.

% What has already been done?\\
% What exactly do we do? Which systems did we study?\\
Among the variety of available methods to calculate quantum transport 
we have chosen the Recursion-Transfer-Matrix (RTM) method
\cite{Hirose1994_150,Brandbyge1997_2637}.
Compared to the Green's function methods which are widely used for calculating
the conductance of nanocontacts the RTM method has the advantage that it 
needs just a potential as input and uses no atomic orbitals. This feature
made it possible to study simple potentials, only determined by geometry
where parameters can be changed easily.

% Define G$_0$\\
% What is the new and interesting point about our research?\\
%- Organization of the paper\\
This paper is organized as follows: In Sect.~\ref{secmeth} we briefly discuss 
the method.  
The results are discussed in Sect.~\ref{secres}. First we examine any 
interference effects when two nano\-con\-strictions are in 
close contact. Then we investigate how the conductance of a torus depends
on its geometry. Finally we present results for the conductance of
the concentric spheres for two cases:
in the first case the constriction has 
cylindrical symmetry and in the second case the symmetry is broken. 
Sect.~\ref{secconc} concludes the paper.
%
%%%%% Method
%
\section{Method}\label{secmeth}
We calculated the conductance using the RTM
method \cite{Brandbyge1997_2637,Hirose1994_150}.
In this method the Schr\"odinger 
equation is solved, where the potential is given on a grid
and the energy $E$ of the transmitting electron is chosen. 
As an ansatz for a constriction along the $z$-direction one expands 
the wave function into plane waves in the $xy$-plane
\begin{equation}
  \psi_{j}(\vec{r}_{\perp},z) = e^{i\vec{k}_{\perp}\cdot \vec{r}_{\perp}}
    \sum_i \phi_{ij}(z) \, e^{i\vec{G}^i_{\perp}\cdot \vec{r}_{\perp}} 
    \label{ansatz}
\end{equation}
where $\vec{G}^i_{\perp}$ are reciprocal lattice vectors in the $xy$-plane
and for the boundary condition one uses 
\begin{eqnarray}
  \phi_{ij}^{\mbox{entr}}(z) & = & \delta_{ij}\; e^{ik^j_zz} + r_{ij}\; e^{-ik^i_zz}\\
    \phi_{ij}^{\mbox{exit}}(z) & = & t_{ij}\; e^{ik^i_zz} 
\end{eqnarray}
for extremal values of $z$, where $\phi_{ij}^{\mbox{entr}}(z)$ denotes the
wave function in the entrance and $\phi_{ij}^{\mbox{exit}}(z)$ denotes the 
wave function in the exit electrode. The components of the reflexion matrix
are denoted by $r_{ij}$.
After the calculation of the transmission matrix 
$\underline{\underline{t}}$ ($(\underline{\underline{t}})_{ij}=t_{ij}$),
one uses the Landauer-B\"uttiker formula
\begin{equation}
  G = \frac{2e^2}{h} \sum_{ij} |t_{ij}|^2 \label{goal}
\end{equation}
to calculate the conductance $G$, where $G$ is measured in units of 
G$_0=2e^2/h$. The eigenchannels are the 
eigenvalues of $\underline{\underline{t}}^{\dagger}\underline{\underline{t}}$
\cite{Brandbyge1997_14956}.

The convergence of the results was carefully tested with respect to the 
energy cut-off, the number of grid points and the distance between 
the grid points. We use atomic units throughout the paper.

%Additionally convergence tests for parameters like the cutoff-energy 
%have been done.

%
%%%%% Results and Discussion
%
\section{Results and Discussion}\label{secres}
\subsection{Two constrictions in close contact}
\begin{figure}
  \includegraphics[scale=0.33]{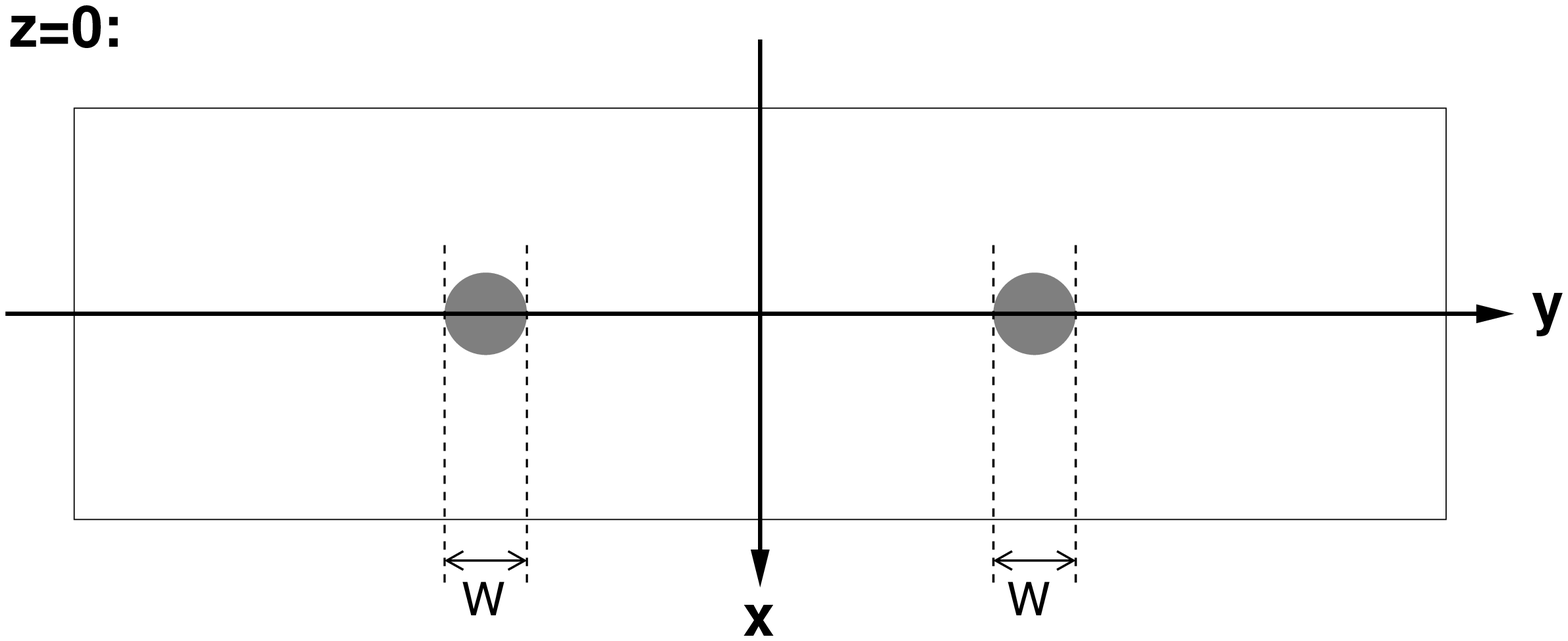}\\
  \vspace{0.5cm}
  \includegraphics[scale=0.33]{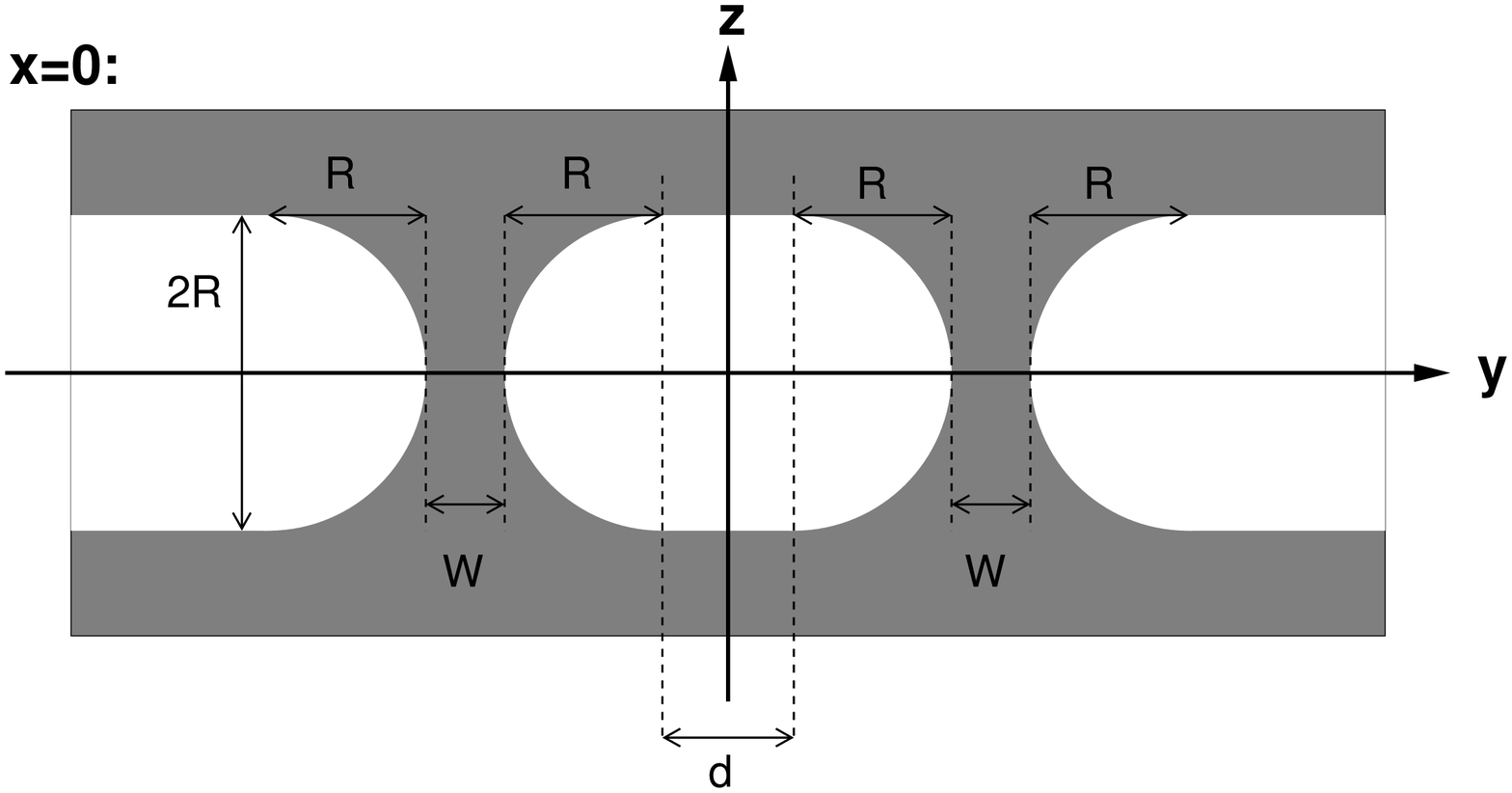}
  \caption{Geometry of the unit cell. The upper panel shows a cut through
	  the contact perpendicular to the $z$-axis for $z=0$. The lower 
	  panel shows a cut through the contact perpendicular to the $x$-axis
	  for $x=0$. At the white area the potential has the value $V=V_0$ and 
	  the grey area denotes zero potential. The electrodes (in which 
	  $V=0$) are located at $z>R$ and $z<-R$.}
  \label{geo2constr}
\end{figure}

To investigate if
two constrictions in close contact interact with each other
we calculated the conductance for the following geometry 
(see Fig.\ \ref{geo2constr}):
\begin{eqnarray}
  V & = & V_0 \; \Theta(R+z) \; \Theta(R-z) \times \nonumber \\
    &   &  \times \; \Theta(x^2+(y+(R+W/2+d/2))^2- \nonumber \\
    &   &  -(R+W/2-\sqrt{R^2-z^2})^2) \times \nonumber \\
    &   &  \times \; \Theta(x^2+(y-(R+W/2+d/2))^2- \nonumber \\
    &   &  -(R+W/2-\sqrt{R^2-z^2})^2)
\end{eqnarray}
We used $V_0=0.404$, $R=5$ and $W=10$. 
The distance between
the constrictions was varied by varying $d$. We probed the contact at an 
energy $E=0.202$.
Our results show that there is no interaction between the two 
constrictions. Additionally we made 
calculations using $R=2$ which are showing no interaction 
either. Due to the geometrical construction of the constriction it is not
possible to calculate distances between the thinnest points of the 
constriction which are smaller than $2R+W$. Anyway for smaller distances 
there may be chemical bonding if real atoms are considered.
%%%%%%%%%% Torus %%%%%%%%%%
\subsection{Torus connected to electrodes}
\begin{figure}
  \includegraphics[scale=0.4]{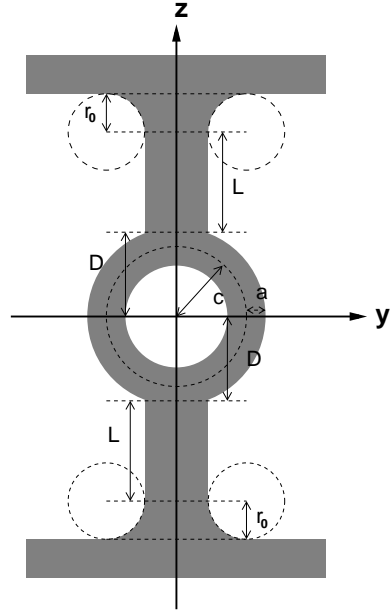}
  \caption{Geometry of the unit cell at $x=0$. At the white area the potential 
	  has the value $V=V_0$ and
          the grey area denotes zero potential. The electrodes (in which
          $V=0$) are located at $z>D+L+r_0$ and $z<-(D+L+r_0)$.}
  \label{geoTorus}
\end{figure}

Another geometry which we studied is a torus. The geometry we used is 
displayed in Fig.\ \ref{geoTorus} where 
a cut along the $yz$-plane is shown. In addition the figure explains all
the parameters. 
So $c$ denotes the radius from the center of the ring to 
the center of the torus tube. The radius of the latter is called $a$.
The torus is cut at $z=\pm D$ in a plane parallel
to the $xy$-plane to open a pipe for the connection to the electrodes.
This distance $D$ also determines the cross-section of 
the pipe since the pipe is
just a prolongation of the hole which is created by the cut.
The total length of this pipe is the sum of 
$L$ and $r_0$, where in addition $r_0$ is the inverse of the curvature
of the opening from the pipe to the electrode. For these parameters
we have chosen the following values: $a=4$, $c=10$, 
$D=11$, $L=4$, $r_0=1$, $V_0=0.404$ 
and $E=0.202$.

To investigate the dependence of the
conductance on these geometrical parameters, calculations have
been done, where we varied one parameter and left the
other parameters constant: 
\begin{figure}
\includegraphics[scale=0.3]{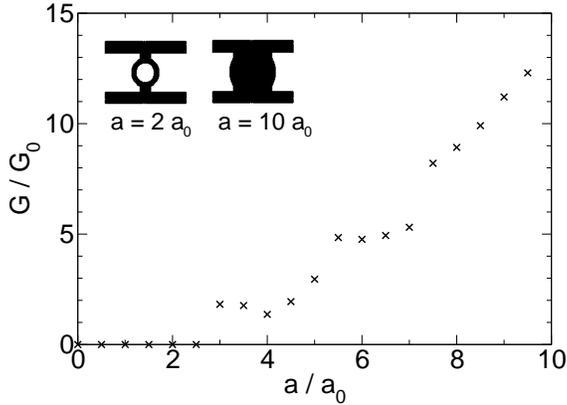}
  \caption{Conductance versus radius of the tube of the torus $a$. 
	  The other parameters have been chosen to be $c=10$, 
	  $D=11$, $L=4$, $r_0=1$, $V_0=0.404$ 
	  and $E=0.202$. The insets
	  show the geometry of the contact perpendicular to the $x$-axis for
	  $a=2$ (left inset) and $a=10$ (right inset).}
  \label{aTorus}
\end{figure}
Figure \ref{aTorus} displays conductance against the radius $a$ of the
tube of the torus. 
The larger $a$ the larger is the conductance. Increasing $a$
increases also the diameter of the pipes, which connect the torus to the 
electrodes, since $D$ is kept constant. 

One can see conductance steps at $a=3$, $a=5.5$ and $a=7.5$.
This can be understood as follows:
In the calculation an electronic energy of $0.202$ is used. This
corresponds to a wavelength $\lambda$ of $9.9$. 
When $a$ is smaller than about
$2.7$ then the tube and the pipes are too narrow to support half a
wavelength and thus there is no current going through. For $2.7 < a < 5.1$, 
half a wavelength fits through the constriction and thus we can
see the first conductance step. 
If $5.1 < a < 7.4$, the system can support one wavelength and 
the conductance jumps up to $G=5$ G$_0$. The next conductance step takes 
place at $a=7.4$ because from $a=7.4$ on until
$a=9.9$ the constriction is able to support $1.5$ wavelengths. If 
$a$ is larger than $9.9$ two wavelengths are fitting inside the system
and so one would expect the next conductance step at $a=10$.

\begin{figure}
\includegraphics[scale=0.3]{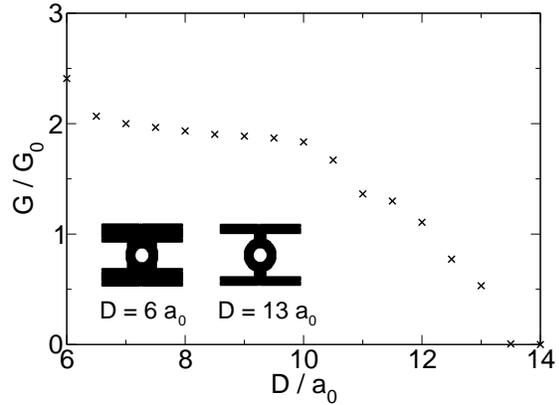}
  \caption{Conductance versus $D$, where $D$ is a measure not only where the 
	  torus is cut but also for the size of the pipes connecting the 
	  torus to the electrodes. The insets show the geometry of the 
	  contact perpendicular to the $x$-axis for $D=6$ (left inset)
	  and $D=13$ (right inset). The other parameters have been
	  chosen to be $a=4$, $c=10$, $L=4$, $r_0=1$,
	  $V_0=0.404$ and $E=0.202$.}
  \label{dTorus}
\end{figure}
The dependence of the conductance on $D$ is displayed in Fig.\
\ref{dTorus}. 
As expected the conductance decreases with a
decreasing cross-section of the pipe (increasing $D$).

\begin{figure}
\includegraphics[scale=0.3]{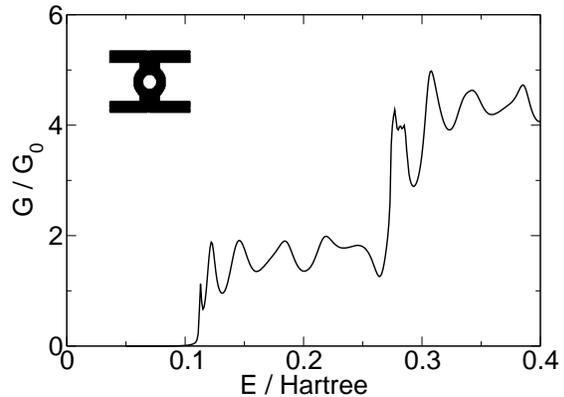}
  \caption{Conductance versus energy $E$ for a geometry where long pipes
	  connecting the torus to the electrodes. The inset shows the 
	  geometry of the contact perpendicular to the $x$-axis. For this
	  calculation the following parameters have been chosen: $a=4$,
	  $c=10$, $D=11$, $L=4$, $r_0=1$ and 
	  $V_0=0.404$.}
  \label{f2Torus}
\end{figure}
Fig.\ \ref{f2Torus} displays the energy-dependence of the conductance for
the considered geometry. For energies larger than about $0.1$ 
the system 
conducts. The increase of the conductance with increasing energy goes 
stepwise. The second conductance step takes place around $E=0.3$.
At the conductance plateaus the conductance shows oscillations. At the 
first conductance plateau the conductance oscillates between $1$ and 
$2$ G$_0$, on the second plateau the conductance takes values between
$4$ and $5$ G$_0$.

To understand the conductance steps one can argue in a similar way than 
when varying $a$ (Fig.\ \ref{aTorus}). In this case $a$ is fixed to
$4$, which means that the wavelength has to be shorter than $16$
($8$) in order to make half a (one) wavelength fit inside 
the tube. The connection to the electrodes has an extension of $7.75$
in $x$-direction and an extension of $16.61$ in $y$-direction.
Converting these lengths in energy yields that the system should start
to be conducting around $E=0.1$ and the next conductance
channel should open around $E=0.3$. 

The oscillations are an effect of the long pipes between the 
electrodes and the torus. Refs. \cite{Szafer1989_300} 
and \cite{VanderMarel1989_7811} calculate
narrow pipes and find similar oscillations the longer the pipes.

\begin{figure}
\includegraphics[scale=0.3]{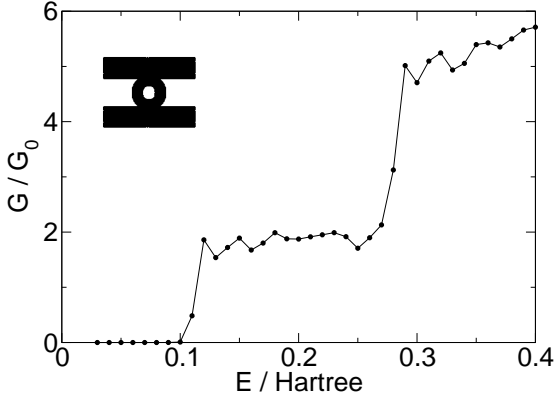}
  \caption{Conductance versus energy $E$ for a geometry where the torus is 
          almost directly connected to the electrodes. The inset shows the
          geometry of the contact perpendicular to the $x$-axis. For this
          calculation the following parameters have been chosen: $a=4$,
          $c=9$, $D=10$, $L=0$, $r_0=0.5$ and
          $V_0=0.404$.}
  \label{f1Torus}
\end{figure}
Figure \ref{f1Torus} shows conductance versus energy for another 
set of parameters. Here the parameters are $a=4$, $c=9$, 
$D=10$, $L=0$, $r_0=0.5$ and $V_0=0.404$.
This means that now the total length
of each pipe is one order of magnitude smaller than in Fig.\ \ref{f2Torus}.
The differences in the other parameters are small, which means that
the main difference is due to the length of the pipes which connect 
the torus to the electrodes. 

We observe that the conductance steps take place at the same energies
than in Fig.\ \ref{f2Torus}. This is due to the same size of the tubes.
In both cases $a=4$ and the extension of the pipe connecting the
torus to the electrodes in $x$-direction is in both cases $7.75$.
Only the extension of these pipes in $y$-direction is here broader
($17.32$ instead of $16.61$).
But this time there are much less 
oscillations at the plateau with a much smaller amplitude. This 
supports the theory that the oscillations may be due to the length of 
the straight pipes.

Though our calculations are only a coarse model, we can nevertheless 
compare them with experiment. There
the conductance of a benzene-1,4,-dithiolate molecule
between gold electrodes has been measured \cite{Reed1997_252} and a step-like
structure of $G(V)$ has been found. Since $G(E)$ can be seen as an
approximation to $G(V)$, our results agree qualitatively with experiment.

%%%%%%%%%% Sphere %%%%%%%%%%
\subsection{Sphere connected to electrodes}

\begin{figure}
  \includegraphics[scale=0.4]{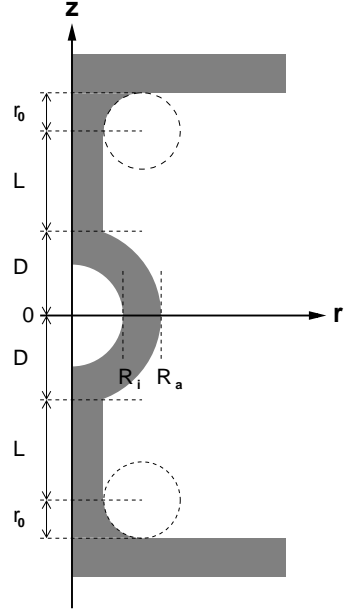}
  \caption{The geometry used to calculate the conductance through a sphere
	  connected to electrodes. Due to the symmetry cylindrical coordinates
	  have been used.
	  At the white area the potential has the value $V=V_0$ and
          the grey area denotes zero potential. The electrodes (in which
          $V=0$) are located at $z>D+L+r_0$ and $z<-(D+L+r_0)$.}
  \label{geoSphere}
\end{figure}
The geometry used for calculating the conductance of a sphere is displayed
in Fig.\ \ref{geoSphere}. Due to the symmetry cylindrical coordinates have 
been used where $r=\sqrt{x^2+y^2}$. At the white area the potential has the 
value $V=V_0$ and the grey area denotes zero potential. Two concentric 
spheres with radii $R_i$ (inner sphere) and $R_a$ (outer sphere) have been 
used in order to be able to change the thickness of the spherical layer
where the electrons are allowed to flow. At a distance $D$ from the center
of the spheres in $\pm z$-direction the sphere is cut parallel to the 
$xy$-plane. This cut determines the radius of the pipes which connects
the sphere to the upper and lower electrode via
\begin{equation}
	r_D = \sqrt{R_a^2 - D^2}. \label{rD}
\end{equation} 
The total length of each pipe is $L+r_0$ where the inverse of $r_0$ is
denoting the curvature of the opening from the pipe to the electrode.

\begin{figure}
\includegraphics[scale=0.40]{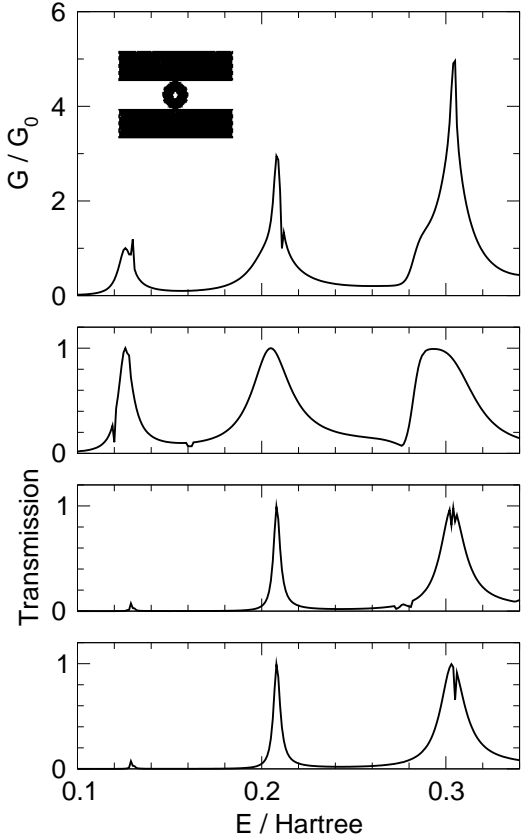}
  \caption{Top panel: Conductance versus energy for a sphere 
	  which is weakly coupled
	  to the electrodes. The three lower panels show the 
	  first three eigenchannels.
	  A cut of the potential along the $yz$-plane is shown in the inset.
	  The parameters used are $R_i=1.7$ and $R_a=7.7$, $D=7.6$, 
	  $L=0.0$ and $r_0=0.5$.}
  \label{f5Sphere}
\end{figure}
The first panel of Fig.\ \ref{f5Sphere} shows the conductance plotted 
against energy for a sphere
which is weakly coupled to the electrodes. In this case "weak coupling"
means that there is only a small connection between the sphere and the 
electrodes. The parameters used in this plot are $R_i=1.7$ and $R_a=7.7$
for the radii of the concentric spheres, $D=7.6$, $L=0.0$ 
and $r_0=0.5$. Thus there is only a small hole of radius $r_D=1.2$ 
connecting the sphere to the electrodes. 

For conduction this small hole must be able to contain at least half a 
wavelength. Using 
\begin{equation}
	E = \frac{2\pi^2}{\lambda^2}
\end{equation}
for conversion between wavelength and energy, the system should start to 
conduct at $0.86$. This energy is out of the range displayed in
Fig.\ \ref{f5Sphere}. But what is seen in Fig.\ \ref{f5Sphere} are 
resonances at special energies. These resonances display the eigenstates
of the sphere.

The three lower panels of Fig.\ \ref{f5Sphere} show the first three 
eigenchannels of the eigenchannel analysis for the geometry used in 
Fig.\ \ref{f5Sphere}. The first
channel is non-degenerate but the second and third eigenchannels are 
degenerate due to the cylindrical symmetry of the constriction.

\begin{figure}
\includegraphics[scale=0.40]{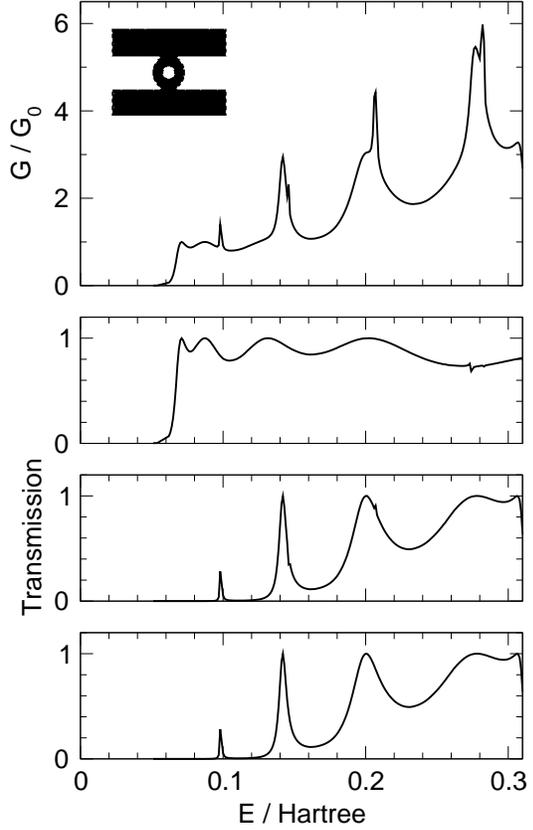}
  \caption{Top panel: Conductance versus energy for a sphere 
	  which is strongly coupled to the electrodes. 
	  The three lower panels show the first three eigenchannels.
	  A cut of the potential along the $yz$-plane
          is shown in the inset. Geometrical parameters used are
	  $R_i=3.7$, $R_a=9.7$, $D=8.6$, $L=0.0$ 
	  and $r_0=1.0$.}
  \label{f2Sphere}
\end{figure}
A different geometry has been used to calculate the conductance versus 
energy curve shown in the first panel of Fig.\ \ref{f2Sphere}. 
Here $R_i=3.7$, $R_a=9.7$, $D=8.6$, $L=0.0$ and $r_0=1.0$. Now $r_D=4.5$ 
and using the same argumentation than above one expects that the system 
conducts for energies larger than $0.06$. Fig.\ 
\ref{f2Sphere} confirms our prediction: around $E=0.06$ the 
conductance jumps from $0$ G$_0$ to $1$ G$_0$. With increasing energy the
conductance increases but sharp peaks are superpositioned on this 
increasing conductance. These sharp peaks again display the eigenstates
of the sphere. An eigenchannel analysis has been made and the first three
eigenchannels are displayed in the lower panels of Fig.\ \ref{f2Sphere}.
The first eigenchannel opens at $0.06$ and stays open for higher 
energies showing some oscillations. The second and the third one are 
degenerate and are showing some resonances before they open around an energy
of $0.2$.

\begin{figure}
\includegraphics[scale=0.40]{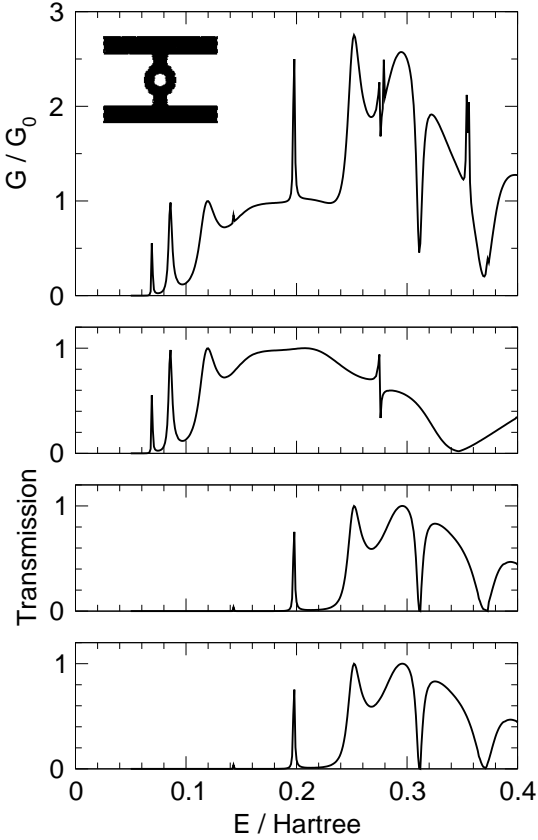}
  \caption{Top panel: Conductance versus energy for a sphere which 
	  is strongly coupled through long pipes to the electrodes. The three
  	  lower panels show the first three eigenchannels. 
	  A cut of the potential along the $yz$-plane is shown in the inset.
	  Geometrical parameters used are $R_i=3.7$, $R_a=9.7$, 
	  $D=8.6$, $L=5.0$ and $r_0=1.0$.}
  \label{f1Sphere}
\end{figure}
The geometries used for obtaining the results in Fig.\ \ref{f5Sphere} and
Fig.\ \ref{f2Sphere} have almost no pipe, the spheres touch more or less
directly the electrodes. What changes if the connection to the electrodes is
made by long pipes? Fig.\ \ref{f1Sphere} shows results obtained for a geometry
where all parameters are the same than in Fig.\ \ref{f2Sphere} except $L$.
This time $L=5.0$ is used instead of $L=0.0$. Now the picture
looks much more complex. $G(E)$ (first panel) shows resonances 
and antiresonances superposing a "background" conductance. 
The complexity may arise due to the much
more complex geometry since every change of geometry (from electrode to pipe,
from pipe to sphere and so on) gives rise to reflections. The three lower
panels of Fig.\ 
\ref{f1Sphere} show the first three eigenchannels of which two 
are degenerate. This time the second and third eigenchannel open at 
an energy of about $0.24$. The opening of these two channels 
can be seen in the first panel of Fig.\ \ref{f1Sphere} 
when the average conductance rises above one conductance quantum.

One can make a qualitative comparison to experiments where the 
conductance of a C$_{60}$-molecule has been measured \cite{Park2000_57}.
In these measurements a peaked structure of the conductance has been found
as well.

%%%%%%%%%% Sphere connected asymmetrically to the electrodes %%%%%%%%%%
\subsection{Sphere connected asymmetrically to the electrodes}
Until now cylindrical
symmetry has been assumed for the constriction. 
In addition all constrictions had a mirror
plane in the $xy$-plane. How does it affect the conductance if these 
symmetries are broken?

\begin{figure}
  \includegraphics[scale=0.4]{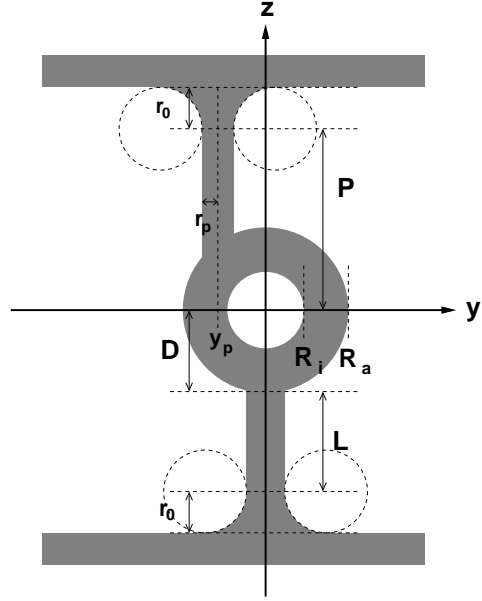}
  \caption{The geometry used to calculate the conductance through a sphere
          connected asymmetrically to electrodes. 
          At the white area the potential has the value $V=V_0$ and
          the grey area denotes zero potential. The electrodes (in which
          $V=0$) are located at $z>P+r_0$ and $z<-(D+L+r_0)$.}
  \label{geoasymSphere}
\end{figure}

To answer these questions calculations have been done where one pipe was 
shifted away from the $z$-axis, where $y_p$ denotes the distance from the
$z$-axis and $x_p=0$ without loss of generality. Fig.\ \ref{geoasymSphere}
shows a cut of the geometry along the $yz$-plane.

\begin{figure}
  \includegraphics[scale=0.3]{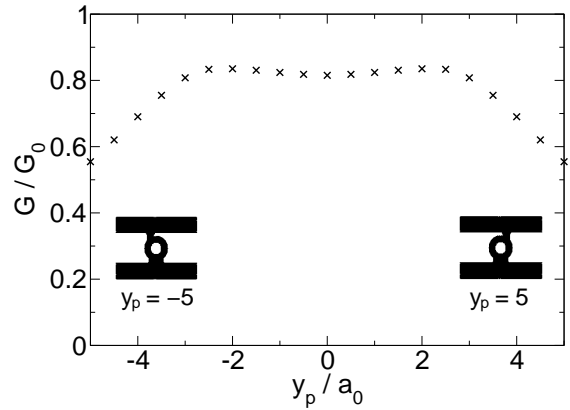}
  \caption{Conductance versus $y_p$. The other parameters take the following
	  values: $R_a=10$, $R_i=5$, $P=11$, $L=2$, 
	  $D=8$, $r_p=3$, $r_0=2$, $V_0=0.404$, 
	  $E=0.202$.}
  \label{yp}
\end{figure}
Fig.\ \ref{yp} shows conductance versus $y_p$. The other parameters were 
chosen to be $R_a=10$, $R_i=5$, $P=11$, $L=2$,
$D=8$, $r_p=3$, $r_0=2$, $V_0=0.404$,
$E=0.202$. If the upper pipe is situated at the $z$-axis 
($y_p=0$), the conductance is about $0.82$ G$_0$. If the upper pipe
is shifted away from the $z$-axis, the conductance increases until it 
reaches a maximum at $y_p=2$. There the conductance is $0.84$ G$_0$.
$y_p=2$ is special in that way, that at this geometry the outer edge
of the pipe has the same value of the $y$-coordinate than the surface of the
inner sphere.
If the distance of the upper pipe to the $z$-axis is increased further,
the conductance decreases.

%
%%%%% Conclusions
%
\section{Conclusions}\label{secconc}
We calculated the conductance of nanocontacts with different geometries
using the recursion-transfer-matrix method. %\cite{Brandbyge1997_2637}
The calculated geometries are two constrictions in close contact, a
torus and two concentric spheres. Our results show that 
geometry matters. It is not only the diameter of the smallest 
part of the constriction, which determines the conductance.
Furthermore small changes in the geometry can have
a large impact on conductance. 

For the case of two constrictions in close contact we found, that
they do not interact with each other.

In the case of the torus we found a steplike behaviour of the
conductance. Each time when there fits half a wavelength more 
inside the tube by either increasing the radius of the tube 
or the energy of the electrons there is a conductance-step.

For the sphere a discrete set of 
energies -- its eigenstates -- is obtained if the
coupling to the electrodes is weak. In the case of strong
coupling the resonances remain but the
overall behaviour resembles more a step structure. 

Moreover the way how the torus or the tube is connected to
the electrodes is important for the conductance.
Connecting the torus or the sphere with pipes to the electrodes
causes more fluctuations in the conductance since the electrons
are suffering more reflections. The total conductance is 
decreased compared to the case with pipes of negligible length.
%
%%%%% Acknowledgements
%
\begin{acknowledgement}
We would like to thank Matti Manninen for fruitful discussions
and the Academy of Finland for financial support.
\end{acknowledgement}
%
%%%%% References
%
%\bibliography{/home/dkoudela/papers/references/references}{}
%\bibliographystyle{daniela}

\end{document}